\documentstyle[aps]{revtex}
\begin{document}


\title{Wigner Functional Approach to Quantum Field Dynamics}

\author{Stanis\l aw Mr\' owczy\' nski\footnote{E-mail:
 MROW@PLEARN.BITNET}}

\address{High-Energy Department, So\l tan Institute for Nuclear Studies,\\
ul. Ho\. za 69, PL - 00-681 Warsaw, Poland}

\author{Berndt M\" uller\footnote{E-mail: MULLER@PHYSICS.PHY.DUKE.EDU}}

\address{Physics Department, Duke University, Durham, N. C. 27708-0305, USA}

\date{29 April 1994}

\maketitle

\begin{abstract}

We introduce the Wigner functional representing a quantum field
in terms of the field amplitudes and their conjugate momenta.  The
equation of motion for the functional of a scalar field point out the
relevance of solutions of the classical field equations to the time
evolution of the quantum field. We discuss the field in thermodynamical
equilibrium and find the explicit solution of the equations of motion
for the so-called ``roll-over'' phase transition.  Finally, we briefly
discuss the approximate methods for the evaluation of the Wigner
functional that may be used to numerically simulate the initial value
problem.

\end{abstract}

\pacs{03.65.Sq, 05.60.+w, 11.10.Ef}

\section{Introduction}

Dynamical processes in relativistic quantum fields are usually described
in terms of the particle excitations of these fields. However, there
are situations where more appropriate degrees of freedom are rather
the field amplitudes themselves and their conjugate momenta.
This happens when as a first approximation the dynamics is well described by
the classical field equations. A well known example here is the evolution of
Higgs fields in the early universe \cite{1,2}, which acquire a finite
vacuum expectation value when the temperature falls below the critical
temperature for the symmetry breaking phase transition. The mechanism
which drives the Higgs field from the symmetric phase, where the vacuum
expectation value vanishes, to the asymmetric phase with a non-zero
expectation value is essentially classical. Such phase transitions have
attracted much attention in the context of inflationary cosmological
models \cite{3}, models of galaxy formation \cite{4}, the creation of the
cosmic baryon number asymmetry during the electroweak phase transition
\cite{5}, and most recently due to the possibility of creating the
so-called misaligned chiral condensates in ultrarelativistic heavy-ion
collisions \cite{6}.

All these situations have in common that one must study the evolution of
a quantum field far off thermal equilibrium, when the initial state of the
quantum field is specified. Most theoretical approaches to this initial
value problem for quantum fields have been based on dynamical equations
for the field expectation values and the Gaussian fluctuations around
those \cite{4,6,7}, assuming that fluctuations around the most probable
path remain small. While this assumption may be correct in certain instances,
it is a well known fact that fluctuations usually do not remain small in
dynamical phase transitions, where domain formation and clustering are
regularly occurring phenomena.

In this paper we present a formalism which appears to be very convenient to
study the evolution of quantum fields from an initial state and which goes
beyond the classical description. The central object of our approach is the
Wigner function,{\it  Wigner functional} in fact, which is the quantum analog
of the classical distribution function in a phase space. Here the phase space,
however, is spanned not by the particle coordinates and momenta, but by the
field amplitudes and their conjugate momenta. The great advantage of the
Wigner function formalism is that, while being fully quantum mechanical,
it remains close in its spirit to the classical description, and that the
classical limit can always be easily obtained. On the other hand, the
simultaneous presence of many ``quasiclassical'' field configurations can
be described with ease, and the fluctuations around these do not have to
remain Gaussian.

An additional motivation for our work derives from the fact
that there are many interesting investigations of nonlinear classical
field equations in Minkowski space.  One can mention here studies
of the chaotic behaviour of classical, in particular, Yang-Mills
fields \cite{8}, searches for exact solutions of the equations of motion
\cite{9} (see e.g. \cite{10} for recent developments), or very specific
predictions concerning multiparticle production in high-energy collisions
\cite{11}.  Whereas the relevance of classical solutions in euclidean space
is safely grounded because they minimize the euclidean action, as in
the case of instantons \cite{9}, no such a argument exists for solutions
in Minkowski space. Attempts to understand the quantum nature of
classical fields have been based on the coherent state representation
of Fock space \cite{12} or semiclassical methods \cite{13}.  For
recent work in this direction, see \cite{14}.

In order to avoid unnecessary complications we restrict considerations
in this paper to scalar quantum fields with quartic self-interaction.
After discussing the definition of the Wigner functional we show that the
equations of motion have the familiar form of a transport equation in phase
space with quantum corrections. We then derive the Wigner functional for a
free scalar field in thermal equilibrium. We discuss how the classical
phase space distribution is obtained in the high temperature limit, and
we obtain the two-point correlation function. We analyze the
``rollover'' of the scalar field in a second-order phase transition, where
the power of the Wigner functional approach becomes apparent, allowing for
the evolution of the quantum field along many simultaneous classical paths.
We briefly discuss the mean field and semiclassical approximation, and
finally we conclude with a suggestion how to obtain
numerical solutions to the initial value problem for quantum fields in
more complex situations.

\section{Wigner Functional}

We consider a real scalar quantum field with the standard
Lagrangian density
\begin{equation}
\hat{\cal L}(t,x) = {1 \over 2} \partial ^{\mu}\hat{\Phi}(t,x)\partial _{\mu}
\hat{\Phi} (t,x)-  {1\over 2} m^2 \hat{\Phi}^2(t,x) + \hat{\cal L}_I(t,x) \;.
\label{2.1}
\end{equation}
The interaction Lagrangian $\hat{\cal L}_I$ is assumed to be a polynomial
in $\Phi$ such as $\hat{\cal L}_I=-{\lambda \over 4!}\Phi ^4$. We use hats
to distinguish operators from corresponding $c-$numbers. For simplicity
most formulas are written for the $1+1$ dimensional field theory, but the
generalization to higher dimensions is straightforward.

Before defining the Wigner functional of the field, let us write down the well
known definition of the single particle Wigner function as
\cite{14a,14b,14c,14d}
\begin{equation}
W(q,p;t) = \int du \; e^{-ipu}
\langle q + {1\over 2}u \vert \; \hat{\rho}(t) \;
\vert q - {1\over 2}u\rangle \;,
\label{2.2}
\end{equation}
where $p$ and $q$ are the particle momentum and position, respectively, while
$\hat{\rho}$ is the time dependent density matrix operator in the
Schr\" odinger picture.
In analogy to (\ref{2.2}) we define the {\it Wigner functional} as
\begin{eqnarray}
W[\Phi (x), \Pi (x); t] =
\int && {\cal D}\phi'(x) \; {\rm exp}\left[ -i\int dx \;\Pi (x) \phi'(x)
\right] \nonumber \\
\times && \langle \Phi (x) + {1 \over 2} \phi'(x) \vert \; \hat{\rho}(t) \;
\vert \Phi(x) - {1 \over 2} \phi'(x) \rangle \;.
\label{2.3}
\end{eqnarray}

It sometimes appears easier to compute the Wigner functional of the fields
in momentum space. However, we then face a little complication.
The field $\Phi (x)$ is real while $\Phi (p)$ is complex, but
the real and imaginary components of $\Phi (p)$ are not
independent from each other because of the constraint $\Phi (-p) = \Phi ^*(p)$.
Thus, we  adopt the following procedure. The real and imaginary parts of $\Phi
(p)$
are treated as independent variables, but
$p \in (0,\infty)$ instead of $p \in (-\infty , \infty )$.
The Wigner functional is then defined as
\begin{eqnarray}
\widetilde W[\Phi (p), \Pi (p); t] =
\int &\;&{\cal D}\phi'(p)\; {\rm exp}\left[ -i \int^{\infty}_0 dp \;
\big( \Pi ^*(p) \phi'(p)    + \Pi (p) \Phi^{\prime *}(p) \big) \right]
\nonumber \\
&\;&\times \langle \Phi (p) + {1 \over 2} \phi'(p) \vert \; \hat{\rho}(t) \;
\vert \Phi(p) - {1 \over 2} \phi'(p) \rangle \;,
\label{2.4}
\end{eqnarray}
with the functional integrations running over real and imaginary components
of $\phi'(p)$.
The transformation from (\ref{2.3}) to (\ref{2.4}) involves a nontrivial
Jacobian
\begin{equation}
{\rm det}\Bigg[ {\delta \Phi (x) \over \delta \Phi (p)}\Bigg] \;.
\label{2.5}
\end{equation}
However, the determinant does not depend on the fields because the Fourier
transformation is a linear mapping; consequently, the Jacobian
affects only the normalization of the Wigner functional.

To clarify the physical meaning of the Wigner functional, we consider the
expectation value of an operator ${\cal O}(\hat{\Phi}, \hat{\Pi} )$.
The expectation value is defined as
\begin{equation}
\langle {\cal O}(\hat{\Phi},\hat{\Pi}) \rangle =
{1 \over Z} \; {\rm Tr}\big[\hat{\rho} (t) \;
{\cal O}(\hat{\Phi},\hat{\Pi}) \big]  \;, \label{2.6}
\end{equation}
with \begin{equation}
Z \equiv {\rm Tr}\hat{\rho}
= \int {\cal D}\Phi(x) {{\cal D}\Pi(x) \over 2\pi} \; W[\Phi,\Pi;t] \;.
\label{2.8}
\end{equation}
As is shown in Appendix A,
\begin{equation}
\langle {\cal O}(\hat{\Phi},\hat{\Pi}) \rangle =
\langle {\cal O}(\Phi ,\Pi ) \rangle \equiv {1 \over Z}
\int {\cal D}\Phi(x) {{\cal D}\Pi(x) \over 2\pi} \;
{\cal O}(\Phi , \Pi ) \; W[\Phi,\Pi;t] \;,
\label{2.7}
\end{equation}
provided that the noncommuting operators in ${\cal O}$ are properly
symmetrized. Operators corresponding to an asymmetric ordering of
the operators $\hat\Phi$ and $\hat\Pi$ must be explicitly expressed
as sums of symmetrized terms.

Equation (\ref{2.7}) shows that the Wigner functional has the same
role in quantum field theory that the density distribution in
the phase space spanned by $\Phi$ and  $\Pi$ has in the classical
field theory. We emphasize, however, that the Wigner functional
is not always positive definite and thus cannot be interpreted as a
probability density in phase space. The analytically tractable cases
discussed below, when the Wigner functional is of Gaussian form in
the variables $\Phi$ and $\Pi$, are somewhat exceptional in this
respect. Of course, the limited classical interpretability of the
Wigner functional does not diminish its usefulness as a representation
of the quantum mechanical density matrix with a simple classical limit.
This will become obvious in the next section, where we study the
relevance of solutions of the classical field equations for the time
evolution of the Wigner functional.

\section{Equation of motion}

The density matrix operator satisfies the equation of motion
\begin{equation}
i\hbar{\partial \over \partial t} \hat{\rho}(t) = [\hat H, \hat{\rho}(t)] \;.
\label{3.1}
\end{equation}
In this section we refrain from setting $\hbar=1$, because we want to discuss
the classical limit of the evolution equation. The Hamiltonian ${\hat H}$
for the real scalar field is
\begin{equation}
\hat H = {1\over 2}\int dx \Big( \hat{\Pi}^2(x) +
\big(\nabla \hat{\Phi} (x) \big) ^2 +
m^2 \hat{\Phi}^2 (x) - \hat{\cal L}_I(x) \Big) \;,
\label{3.2}
\end{equation}
where $\hat{\Pi} \equiv \delta \hat{\cal L}/\delta \dot{\hat{\Phi}} =
\dot{\hat{\Phi}}$ denotes the conjugate momentum operator.

As discussed in Appendix B, one derives from Eq. (\ref{3.1}) the following
equation of motion of the Wigner functional (\ref{2.3})
\begin{equation}
\left[ {\partial \over \partial t} + \int dx \; \left( \Pi (x) \;
{\delta \over \delta \Phi (x)} - \big( m^2 \Phi (x) - \nabla^2 \Phi (x) \big)
{\delta \over \delta \Pi (x)} + {\cal K}_I(x) \right) \right]
W[\Phi,\Pi;t] = 0 \;,
\label{3.3}
\end{equation}
where
\begin{equation}
{\cal K}_I(x) \equiv  - {i\over\hbar}{\cal L}_I\left(\Phi (x)
+{i\hbar\over 2}{\delta \over \delta \Pi (x)}\right) +
{i\over\hbar}{\cal L}_I\left(\Phi (x) -{i\hbar\over 2}
{\delta \over \delta \Pi (x)}\right) \;.
\label{3.4}
\end{equation}
For ${\cal L}_I(\Phi) = - {\lambda \over 4!}\Phi ^4$ we find
\begin{equation}
{\cal K}_I = {\lambda \over 6} \left( - \Phi^3 (x) {\delta\over\delta\Pi (x)}
+ {\hbar^2 \over 4} \Phi (x) {\delta^3 \over \delta \Pi^3 (x)} \right) \;.
\label{3.5}
\end{equation}
In particular, the interaction term ${\cal K}_I$ always terminates for
renormalizable quantum field theories in $(3+1)$ space-time dimensions,
since their Lagrangians contain at most quartic interaction terms.

One sees that the equation (\ref{3.3}) has the familiar structure of a
transport equation. When the higher derivative term in the interaction
(\ref{3.5}) is neglected, eq. (\ref{3.3}) can be written in the form of
the classical Liouville equation
\begin{equation}
\left[ {\partial \over \partial t} + \int dx \; \left(
{\delta H \over \delta \Pi (x)} \;
{\delta \over \delta \Phi (x)} -
{\delta H \over \delta \Phi (x)} \;
{\delta \over \delta \Pi (x)} \right) \right]
W[\Phi,\Pi;t] = 0 \;,
\label{3.6}
\end{equation}
where $H$ is  the classical Hamiltonian. The neglected term is proportional
to $\hbar^2$, showing that corrections to the classical phase space
evolution are of second order in $\hbar$. We will later (in section 7)
briefly discuss how the quantum corrections to the time evolution
of the ``classical" Wigner functional could be obtained.

\section{Free Fields in Thermal Equilibrium}

Let us consider the scalar field in thermodynamical equilibrium.
The density operator is then
\begin{equation}
\hat{\rho} = {1\over Z} e^{-\beta \hat{H}} \;,
\label{4.1}
\end{equation}
where $\beta = T^{-1}$ is the inverse temperature and
$Z= {\rm Tr}\, e^{-\beta \hat{H}}$ is the partition function.
Since there is no conserved charge carried by the real scalar field,
there is no chemical potential in Eq. (\ref{4.1}).

It appears easier first to compute the Wigner functional of the fields
in the momentum space (\ref{2.4}) and then transform it into the
coordinate space representation (\ref{2.3}). The Hamiltonian
\begin{equation}
\hat H_0 = \int^{\infty}_0 {dp \over 2\pi}
\left( \hat{\Pi}^{\dagger}(p)\hat{\Pi}(p)
+(p^2+ m^2) \hat{\Phi}^{\dagger}(p) \hat{\Phi}(p) \right)\;
\label{4.2}
\end{equation}
can be identified as the sum of independent harmonic oscillator Hamiltonians
each representing the mode of momentum $p$. Using this analogy one finds,
as shown in Appendix C, the equilibrium Wigner functional as
\begin{equation}
\widetilde W_\beta[\Phi,\Pi] =  C \; {\rm exp}\left[- {1 \over 2}
\beta \int_{-\infty}^\infty {dp \over 2\pi} \;
\tilde{\Delta}_{\beta}(p) \;\Bigl( \Pi^*(p)\Pi (p)
+E^2(p)\Phi^*(p) \Phi (p) \Bigr) \right]\;,
\label{4.3}
\end{equation}
with $E(p) \equiv \sqrt{p^2 + m^2}$, the thermal weight factor
\begin{equation}
\tilde{\Delta}_{\beta}(p)
= {2 \over \beta E(p)} {\rm th} {\beta E(p) \over 2}  ,
\label{4.3a}
\end{equation}
and the divergent normalization factor
\begin{equation}
C \equiv {\rm exp} \left[ \int_{-\infty}^\infty {dp \over 2\pi}\;
{\rm ln \; th}{\beta E(p) \over 2} \right] \;,
\label{4.4}
\end{equation}
which represents the contributions from zero modes.  Since the expressions
under the momentum integrals in (\ref{4.3}) and (\ref{4.4}) are even functions
of momentum, the integrals are extended from $-\infty$ to $\infty$ not from
0 to $\infty$, as in the definition (\ref{2.4}). The Wigner functional
(\ref{4.3}) is normalized to unity.

In the high temperature limit, where $C, \tilde{\Delta}_{\beta}(p) \to 1$,
we reproduce the classical result i.e.
\begin{equation}
\widetilde W_\beta^{\rm cl}[\Phi,\Pi]  =
 {\rm exp}\left[- {1 \over 2} \beta \int {dp \over 2\pi} \;
\Bigl( \Pi^* (p)\Pi (p) +E^2(p)\Phi^* (p) \Phi (p) \Bigr) \right]\;.
\label{4.5}
\end{equation}
In this case the Wigner functional depends only on the total energy of the
system.

One immediately finds from (\ref{4.3}) that
\begin{equation}
\langle \hat{\Phi}^{\dagger}(p) \hat{\Phi} (p) \rangle =
{1 \over 2 \; E(p) \; {\rm th}{\beta E(p) \over 2}}
\;, \qquad \langle \hat{\Pi}^{\dagger}(p) \hat{\Pi} (p) \rangle =
{E(p) \over 2 \; {\rm th}{\beta E(p) \over 2}} \;,
\label{4.6}
\end{equation}
where $\langle ... \rangle$ is given by Eq. (\ref{2.7}). Therefore,
\begin{equation}
\sqrt{ \langle \hat{\Phi}^{\dagger}(p) \hat{\Phi} (p) \rangle \;
\langle \hat{\Pi}^{\dagger}(p) \hat{\Pi} (p) \rangle } =
{1\over 2 \; {\rm th}{\beta E(p) \over 2}} \ge {1\over 2} \; ,
\label{4.7}
\end{equation}
showing that the uncertainty principle is built into the Wigner functional.

Knowing the Wigner functional (\ref{2.4}) one easily finds the Wigner
functional (\ref{2.3}). As mentioned previously, the Jacobian (\ref{2.5})
only modifies the normalization factor, which is infinite anyway. Thus,
\begin{equation}
W_\beta[\Phi(x),\Pi(x)] = C'\;
 {\rm exp} \left[- \beta \int dx \;dx' \; {\cal H}(x,x') \right] \;,
\label{4.8}
\end{equation}
with
\begin{equation}
{\cal H}(x,x') \equiv {1 \over 2}\; \Delta (x-x') \;
 \Bigl( \Pi (x) \Pi (x') + \nabla \Phi (x) \nabla \Phi (x') +
m^2 \Phi (x) \Phi (x') \Bigr) \;,
\label{4.9}
\end{equation}
where
\begin{equation}
\Delta_{\beta} (x) = \int_{-\infty}^\infty {dp \over 2\pi} \; e^{-ipx} \;
\tilde{\Delta}_{\beta}(p)\; .
\label{4.10}
\end{equation}
One sees that $\Delta_{\beta} (x) \approx \delta (x)$ for $m^{-1}, \vert {x}
\vert \gg \beta$ which corresponds to the classical limit.
For $m = 0$ the integral (\ref{4.10}) can be computed analytically
in one dimension \cite{15} as
\begin{equation}
\Delta_{\beta} (x) = {2 \over \beta \pi} \; {\rm ln \; cth}
{\pi \vert x \vert \over 2 \beta} \;,
\label{4.11}
\end{equation}
and approximated in the following way
\begin{equation}
\Delta_{\beta} (x) \approx \cases{ -{ 2 \over \beta \pi} \; {\rm ln} \;
{\vert x \vert \over \beta} & $\;\;\;\;{\rm for} \;\;\;
\vert x \vert \ \ll \beta $ \cr
{ 4 \over \beta \pi} \; {\rm exp} \left(- {\pi \vert x \vert \over \beta}
\right) & $\;\;\;\;{\rm for} \;\;\; \vert x \vert \ \gg \beta \;.$ \cr}
\label{4.12}
\end{equation}
One easily checks that the Wigner functional (\ref{4.8}) represents a
time-independent solution of the equation of motion (\ref{3.3}).

\section{Equilibrium correlation function}

As usually, the correlation function is obtained from the generating
functional,
which for the free fields in equilibrium is defined as
\begin{eqnarray}
{\cal Z}[j(x)] & \equiv &
\int {\cal D}\Phi {{\cal D}\Pi \over 2\pi} \;
W[\Phi,\Pi] \exp \left( \beta \int dx\; \Phi (x) j(x) \right)
\nonumber \\
& = & \int {\cal D}\Phi {{\cal D}\Pi \over 2\pi} \;
{\rm exp} \left[- \beta \int dx \;dx' \; \Bigl( {\cal H}(x,x')
 - \delta (x- x') \Phi (x) j(x') \Bigr) \right] \;,
\label{5.1}
\end{eqnarray}
where ${\cal H}(x,x')$ is given by Eq. (\ref{4.9}) and $j(x)$ denotes the
external classical current.
Due to Eq. (\ref{2.7}) the correlation function can be expressed as
\begin{equation}
\langle \hat{\Phi}(x) \hat{\Phi}(y) \rangle =
{1 \over \beta^2} \; {1 \over {\cal Z}[j]}
{\delta ^2 {\cal Z}[j] \over \delta j(y) \delta j(x)}
\Bigg\vert_{j(x)=0} \;.
\label{5.2}
\end{equation}
As shown in Appendix D, the generating functional is
\begin{equation}
{\cal Z}[j(x)] = {\cal N} \;
 {\rm exp} \left[ {\beta \over 2} \int dx \;dx' \;
j(x) {\cal G}(x-x') j(x') \right]\;,
\label{5.3}
\end{equation}
where ${\cal N}$ is the normalization constant and
\begin{equation}
{\cal G}(x) = \int {dp \over 2\pi} \;
{e^{-ipx} \over \tilde{\Delta}_{\beta}(p) \; \big[ p^2 + m^2 \big] } \;,
\label{5.4}
\end{equation}
with $\tilde{\Delta}_{\beta}(p)$ given in Eq. (\ref{4.3a}).  Thus, the
correlation function (\ref{5.2}) equals
\begin{equation}
\langle \hat{\Phi}(x) \hat{\Phi}(y) \rangle = {\cal G}(x-y) \;.
\label{5.5}
\end{equation}

The integral (\ref{5.4}) can be evaluated analytically \cite{15} in
three dimensions for $m=0$
\begin{equation}
{\cal G}({\bf x}) = {1 \over 4\pi \vert {\bf x} \vert } \;
{ {\rm sh}({2\pi \vert {\bf x} \vert / \beta}) \over
  {\rm ch}({2\pi \vert {\bf x} \vert / \beta}) -1}  \;,
\label{5.6}
\end{equation}
\begin{equation}
{\cal G}({\bf x}) \approx
\cases{{1 \over 4\pi \vert {\bf x} \vert}\;
& for $\;\;\;\;\; \beta \ll \vert {\bf x} \vert $, \cr
{\beta \over 4\pi^2 \vert {\bf x} \vert ^2}\;
& for $\;\;\;\;\; \beta \gg \vert {\bf x} \vert $. \cr}
\label{5.7}
\end{equation}

For a finite mass the integral (\ref{5.4}) can be calculated in an
approximate way and for three dimensions one finds
\begin{equation}
{\cal G}({\bf x}) \approx  {1 \over 4\pi \vert {\bf x} \vert}\;
e^{-m \vert {\bf x} \vert}
\label{5.8a}
\end{equation}
for $m^{-1}, \vert {\bf x} \vert \gg \beta$,
which corresponds to the classical limit, and
\begin{equation}
{\cal G}({\bf x}) \approx  {\beta m \over 4\pi^2 \vert {\bf x} \vert}\;
K_1(m \vert {\bf x} \vert ) \approx
{\beta m^{1/2} \over 2^{5/2}\pi^{3/2} \vert {\bf x} \vert ^{3/2}} \;
e^{-m \vert {\bf x} \vert}  \;,
\label{5.8b}
\end{equation}
where the first approximate equality holds for $m^{-1} \ll \beta $,
while the second one requires additionally $m^{-1} \ll \vert {\bf x} \vert$.

\section{``Rollover" Phase Transition}

Let us consider the scalar field in variable heat bath, undergoing a phase
transition with falling temperature. The mass squared of the field,
effectively being temperature dependent, is initially assumed to be
positive. We are interested in the evolution of the quantum field
with spatially homogeneous initial conditions when,
suddenly, due to a rapid decrease of the temperature, the mass squared
acquires a negative value. This model is an idealization of instabilities
arising in, e.g., inflationary cosmology and during the chiral phase
transition in expanding dense hadronic matter. Thus, we study the time
evolution of the field described by the Hamiltonian
\begin{equation}
\hat H (t) = \int_0^{\infty} {dp \over 2\pi}
\Big( \hat{\Pi}^{\dagger}(p)\hat{\Pi}(p)
+(m^2(t)+p^2)\; \hat{\Phi}^{\dagger}(p) \hat{\Phi}(p) \Big)\;,
\label{6.1}
\end{equation}
with
\begin{equation}
m^2(t) = \cases{m^2 > 0    \;
& for $\;\; t < 0 \;$, \cr
             -\mu^2 < 0 \;
& for $\;\; t > 0 \;$. \cr}
\label{6.2}
\end{equation}

At $t<0$ the state of the system is described by the equilibrium Wigner
functional (\ref{4.3}).  To find the system evolution starting with $t=0$,
we split the Hamiltonian (\ref{6.1}) for $t > 0$ into stable and unstable
modes as
\begin{eqnarray}
\hat H (t>0) &= & \int_0^{\mu} {dp \over 2\pi}
\Big( \hat{\Pi}^{\dagger}(p)\hat{\Pi}(p)
- \omega_-^2(p) \; \hat{\Phi}^{\dagger}(p) \hat{\Phi}(p) \Big)
\nonumber \\
&\quad &+\int_{\mu}^{\infty} {dp \over 2\pi}
\Big( \hat{\Pi}^{\dagger}(p)\hat{\Pi}(p)
+ \omega_+^2(p) \; \hat{\Phi}^{\dagger}(p) \hat{\Phi}(p) \Big)\;,
\label{6.3}
\end{eqnarray}
where $\omega_{\pm}(p) \equiv \sqrt{\pm (p^2 -\mu^2)}$ is a real number.

One sees that the modes are independent from each other, thus we discuss
for a moment a single mode, stable or unstable.  Since the Wigner
functional equation of motion (\ref{3.6}) with $m^2=-\mu^2$ coincides with
the classical Liouville equation, it can be solved in a way which is well
known from classical mechanics. Specifically, the single mode equation of
motion is solved by
\begin{equation}
W(\Phi_p, \Pi_p;t) = W(\Phi_p(-t),\Pi_p(-t);0) \;,
\label{6.4}
\end{equation}
where $\Phi_p(t)$ is the solution of the classical field equation
\begin{equation}
{d^2 \over dt^2} \; \Phi_p(t) \pm \omega_{\pm}^2(p) \; \Phi_p(t) = 0 \;
\label{6.5}
\end{equation}
with $\Pi_p(t) = \dot\Phi_p(t)$ and the initial conditions
$\Phi_p(0)=\Phi_p$, $\Pi_p(0)=\Pi_p$.

Solving Eq. (\ref{6.5}) one finds the single mode Wigner function
(\ref{6.4}), and then gets the whole functional as a product of the single
mode functions. Thus,
\begin{equation}
\widetilde W[\Phi,\Pi;t] =   C \;
{\rm exp}\left[- {1 \over 2} \beta \int_{-\infty}^\infty {dp \over 2\pi} \;
\tilde{\Delta}_{\beta}(p) \;
\Big( \Pi^*_0(p)\Pi_0 (p) +E^2(p)\Phi^*_0(p) \Phi_0 (p) \Big) \right]\;,
\label{6.6}
\end{equation}
where
\begin{eqnarray}
\Phi_0 (p) & = &\Phi (p) \; {\rm ch}(\omega_-(p)t) - {1 \over \omega_-(p)} \;
 \Pi (p) \; {\rm sh}(\omega_-(p)t) \label{6.7a} \\
 \Pi_0 (p) & = &\Pi (p) \; {\rm ch}(\omega_-(p)t) - \omega_-(p) \;
 \Phi (p) \; {\rm sh}(\omega_-(p)t)
\label{6.7b}
\end{eqnarray}
for $0< p < \mu $ and
\begin{eqnarray}
\Phi_0 (p) & = &\Phi (p) \; {\rm cos}(\omega_+(p)t)
- {1 \over \omega_+(p)} \;
\Pi (p) \; {\rm sin}(\omega_+(p)t) \label{6.7c}\; , \\
\Pi_0 (p) & = &\Pi (p) \; {\rm cos}(\omega_+(p)t) + \omega_+(p) \;
\Phi (p) \; {\rm sin}(\omega_+(p)t) \;,
\label{6.7d}
\end{eqnarray}
for $p > \mu $.
It should be stressed that the solution (\ref{6.6}) is exact and fully quantum
mechanical.

If one is interested only in the  $\Phi -$dependence of the Wigner
functional, $\Pi $  is integrated over and one gets
\begin{equation}
\int {{\cal D}\Pi \over 2\pi} \widetilde W[\Phi,\Pi;t] =
C' {\rm exp}\left[- {\beta \over 2} \int_0^{\infty}
{dp \over 2\pi} \; \tilde{\Delta}_{\beta}(p) \;
\chi(p,t) \; \Phi^*(p) \Phi (p) \right]\;,
\label{6.8}
\end{equation}
where
\begin{eqnarray}
\chi(p,t) & = & \Theta (\mu - p) \;
{E^2(p)\omega_-^2(p) \over \omega_-^2(p) \; {\rm ch}^2(\omega_-(p)t) +
E^2(p) \; {\rm sh}^2(\omega_-(p)t) } \nonumber \\
& & + \Theta (p - \mu ) \;
{E^2(p)\omega_+^2(p) \over \omega_+^2(p) \; {\rm cos}^2(\omega_+(p)t) +
E^2(p) \; {\rm sin}^2(\omega_+(p)t) }\;.
\label{6.9}
\end{eqnarray}

Using Eq. (\ref{6.8}) one easily finds how the  unstable modes grow
\begin{equation}
\langle \hat{\Phi}^{\dagger}(p) \hat{\Phi} (p) \rangle_t =
\langle \hat{\Phi}^{\dagger}(p) \hat{\Phi} (p) \rangle_0 \;
\Big( {\rm ch}^2(\omega_-(p)t) + {E^2(p) \over \omega_-^2(p)}
{\rm sh}^2(\omega_-(p)t) \Big) \;,
\label{6.10}
\end{equation}
with
$\langle \hat{\Phi}^{\dagger}(p) \hat{\Phi} (p) \rangle_0$ given by
Eq. (\ref{4.6}).  The unstable mode growth is usually obtained directly
from the classical equation of motion (\ref{6.5}). In such a case however,
one finds only the first term in the r.h.s of Eq. (\ref{6.10}). The
second term, which results from the proper incorporation of the initial
condition and the interplay between $\Phi$ and $\Pi$ in (\ref{6.6}), is
missing.  Then it is argued that the zero-momentum mode grows the fastest.
This statement, however, is not quite correct. Approximating the r.h.s of
Eq. (\ref{6.10}) as
\begin{equation}
\langle \hat{\Phi}^{\dagger}(p) \hat{\Phi} (p) \rangle_t =
\langle \hat{\Phi}^{\dagger}(p) \hat{\Phi} (p) \rangle_0 \;
\Big( 1 + (\mu^2 + m^2 )t^2 + {\cal O}\big(\omega_-^4(p)t^4\big) \Big)\;.
\label{6.11}
\end{equation}
one sees that when the instability initially (for $t\ll\omega_-^{-1}(p)$)
all unstable modes ($p^2 < \mu^2$) grow at the same rate,
and the zero-momentum mode becomes dominant only at a later stage. This
observation might be of physical significance since the  experimental
detection of the misaligned chiral condenstates [6], if they are
produced in heavy-ion collisions, would be difficult if the
phenomenon occurred  only for pions with approximately zero momentum.
Eq. (\ref{6.11}) suggests that the pions with nonzero momentum have
a chance to contribute to the condensate.

The two-point correlation function is also of interest here, because it
contains information about domain growth after the phase transition.
The generating functional ${\cal Z}[j]$ defined in (\ref{5.1}) depends
explicitly on time in the present case. A straightforward calculation
shows that the generating functional again has the form (\ref{5.3}),
where the time-dependent correlation function now is given by
\begin{equation}
{\cal G}(x,t) = \int {dp \over 2\pi} \;
{e^{-ipx} \over \tilde{\Delta}_{\beta}(p) \; \chi(p,t) } \;.
\label{6.12}
\end{equation}
The late-time properties of ${\cal G}(x,t)$ are determined by the
exponentially growing modes with $p^2<\mu^2$, for which $\chi(p,t)
\propto \exp[-2\omega_-(p)t] \to 0$. Expanding $\omega_-(p)$ around
$p=0$ and carrying out the three-dimensional Gaussian integral over
$p$ one finds:
\begin{equation}
{\cal G}(x,t) \; {\buildrel {t\to\infty} \over \longrightarrow} \;
{\beta m\mu \over 64\; {\rm th}\, (\beta m/2)}
\Bigl( 1 + {\mu^2\over m^2} \Bigr) (\pi\mu t)^{-3/2}
\exp \left( 2\mu t - {\mu x^2 \over 4t} \right) \; .
\label{6.13}
\end{equation}
{}From this result one can read off that the size of correlated scalar
field domains grows as $\langle x^2 \rangle \sim 4t/\mu$.

\section{Approximate methods}

There are only a few situations when the exact equation of motion (\ref{3.3})
can be solved analytically.  Thus, there is an obvious need to develop
approximate methods to study problems more complicated than those
discussed above.  We  briefly present in this section three approaches
which seem to be promising.

\subsection{Mean-field approximation}

The mean field approximation, which has been frequently used to study
symmetry breaking phase transitions, is implemented by replacing in
the initial Lagrangian
the terms which are cubic, quartic, etc. in fields by the products of
the fields and their expectation values. In the case of
$\hat{\cal L}_I(\Phi) = - {\lambda \over 4!}\hat{\Phi}^4$ the replacement is
\begin{equation}
{\lambda \over 4!}\hat{\Phi}^4(x) \rightarrow
{\lambda \over 4}\langle \hat{\Phi}^2(x) \rangle \; \hat{\Phi}^2(x) \;,
\label{7.1}
\end{equation}
or equivalently, we replace the mass $m^2$ in the free field equation
of motion by the effective mass
$m_*^2 \equiv m^2 + {\lambda \over 2}\langle \Phi ^2(x) \rangle$.
The combinatorial
factor $4!$ changes into 4 in Eq. (\ref{7.1}) because there are 6 ways to
select 2 fields out of 4. Since $\langle \hat{\Phi}^2(x) \rangle$ is
determined by the functional $W$ via Eq. (\ref{2.7}), we get a
self-consistent nonlinear equation for $W$.

Let us briefly discuss the mean field approximation for the
case of thermodynamical equilibrium. The Wigner functional is then given
by Eq. (\ref{4.3}) derived for the free fields.  However, instead of the
mass there is the effective mass determined by the following gap equation
\begin{equation}
m^2_* = m^2 + {\lambda \over 2 } \; {\cal G}(0) \;,
\label{7.2}
\end{equation}
where ${\cal G}(0)$ is defined by Eq. (\ref{5.4}) with $m_*$ substituting $m$.
Before the equation (\ref{7.2}) is solved one has to regulate ${\cal G}(0)$
subtracting from it a divergent zero temperature contribution. After
this procedure we find the well known gap equation
\begin{equation}
m^2_* = m^2 + {\lambda \over 2 } \int_{-\infty}^{\infty} {dp \over 2\pi}\;
{ \sqrt{p^2 + m^2_*} \over {\rm exp}\big(\beta \sqrt{p^2+m^2_*}\big)  - 1} \;
{1 \over p^2 + m^2_*} \;,
\label{7.3}
\end{equation}
which is further analysed in \cite{2}. It is worth mentioning that the
standard result is reproduced only after the correlation function
is renormalized.

\subsection{Semiclassical approximation}

This approximation might be particularly useful when the system of
interest is, obviously, ``semiclasical" and the ``classical" solution,
analytically or numerically, is known. Then, the quantum correction can
be found in the following way. The equation of motion (\ref{3.3}) is
written down in a symbolic notation as
\begin{equation}
\big( \hat{K}_{\rm c} + \hbar^2 \hat{K}_{\rm q}) \; W = 0\;,
\label{7.4}
\end{equation}
where $\hat{K}_{\rm c}$ and $\hat{K}_{\rm q}$ represent the ``classical"
and ``quantum" part of the operator acting on the Wigner functional,
which is expanded as $W = W_{\rm c} + \hbar^2 W_{\rm q}$.  Since
$\hat{K}_{\rm c} W_{\rm c} = 0$, we find $W_{\rm q}$ as a solution
of the equation
\begin{equation}
\hat{K}_{\rm c} W_{\rm q} = - \hat{K}_{\rm q} W_{\rm c} + {\cal O}(\hbar^2)\;.
\label{7.5}
\end{equation}
For a single particle Wigner function one finds
\begin{equation}
W_{\rm q}(q,p;t) = \int dt' dq' dp' \; G(q,q',p,p';t,t') \;
\hat{K}_{\rm q} W_{\rm c}(q',p';t') \;
\label{7.6}
\end{equation}
where the Green function $G(q,q',p,p';t,t')$ solves the equation
\begin{equation}
\hat{K}_{\rm c} G(q,q',p,p';t,t') = {\partial G \over \partial t} +
{\partial H \over \partial p} {\partial G \over \partial q} -
{\partial H \over \partial q} {\partial G \over \partial p} =
\delta (t-t')\; \delta (q-q') \;  \delta (p-p') \;.
\label{7.7}
\end{equation}
Since solutions of the classical equations of motion are assumed
to be known, the Green
function can be explicitely written as
\begin{equation}
G(q,q',p,p';t,t') = \Theta (t-t') \; \delta (q-\tilde q(t)-q'+\tilde q(t')) \;
\delta (p-\tilde p(t)-p'+\tilde p(t')) \;,
\label{7.8}
\end{equation}
with $(\tilde q(t),\tilde p(t))$ being the classical solution.
The generalization of Eqs. (\ref{7.6}) and (\ref{7.8}) to the case of a
quantum field is trivial, only the notation is more complicated.

\subsection{Numerical Simulation of the Initial Value Problem}

The main goal of this study has been to find a formalism to analyze
temporal evolution of the quantum fields beyond the classical limit.
Here we argue that equation of motion (\ref{3.3}) is indeed numerically
tractable.

The first step is to define the field $\Phi (x)$ on a lattice.
Then, the equation of motion (\ref{3.3}) resembles a many particle
transport equation with every site of the lattice representing
a single particle that interacts with an external polynomial
potential (due to the mass and interaction terms of the Lagrangian)
and with its nearest neighbors (via the gradient term). We can then
adopt, e.g., the method developed by John and Remler \cite{16} to
solve the quantum transport or quantum Liouville equation of a many
particle system. The method is sketched below for a single particle;
a generalization to many particles is  straightforward.

A single particle Wigner function is represented by a finite sum
over discrete points in phase space. Since the Wigner function is
not positive definite, the points may have positive or negative
signature. Then, each point representing a test particle is evolved
in such a way that every step of the ``classical" evolution is followed
by ``quantum" step, which converts a single momentum point into several
ones. Specifically, the delta function in momentum space is represented
by a gaussian distribution centered around the classical position.
The higher momentum gradients, which are responsible for the quantum
evolution, act on this distribution. The resulting function, which has
positive and negative components, is again represented by discrete
points, i.e. by new test particles with positive or negative signature.
To avoid unlimited proliferation of the test particles, those with
closely neighboring trajectories can be allowed to merge.  It was
demonstrated in \cite{16} that the method successfully works provided
the initial Wigner function is sufficiently smooth in phase space.
We intend to apply this method in future work to study the quantum
field evolution.

\section{Summary and outlook}

We have introduced the formalism which describes the quantum field
dynamics in terms of the field amplitudes $\Phi$ and their conjugate
momenta $\Pi$. The Wigner functional playing the central role in our
approach provides a density in a phase space spanned by $\Phi$ and
$\Pi$. We have derived the equation of motion of the functional, which
is of the form of the quantum transport or Liouville equation. It has
been shown in the last section how the knowledge of the classical
solution of this equation can be used to find quantum corrections.
The free fields in thermodynamical equilibrium have been discussed in
detail, and we found an explicit analytical solution of the field
analog of the ``upside-down" harmonic oscillator.  The solution
incorporates in a very natural way the ``thermal" initial conditions,
its physical meaning is intuitive and transparent. In the last section
we briefly discussed the approximate methods which will be needed to
solve more complicated problems.

We intend to apply our approach to study the role of quantum fluctuations
in temporal evolution of the symmetry breaking phase transitions
beyond the gaussian approximation \cite{7}.  As mentioned in the
introduction our formalism also appears useful to investigate the
quantum content of solutions of the classical field equations of motion.

\begin{acknowledgments}

The authors are grateful to C. Gong and S.G. Matinyan for fruitful
discussions, and to David Fairlie for useful comments about the history
of the Wigner function approach to quantum mechanics.
St.M. would like to thank to the Physics Department
of Duke University for its warm hospitality. This work has been partially
supported by the U.S. Department of Energy (Grant No. DE-FG05-90ER40592)
and by the Polish Committee of Scientific Research (Grant No. 20960-91-01).

\end{acknowledgments}

\appendix

\section{}

In this Appendix we prove the equations (\ref{2.7}) and (\ref{2.8}), which
show that the Wigner functional can be treated as a density in the phase
space spanned by $\Phi$ and $\Pi$.  The formula (\ref{2.8}) is obvious if
one observes that
\begin{equation}
\int {{\cal D}\Pi \over 2\pi} \;
{\rm exp}\left[ -i \int dx \; \Pi (x) \phi'(x) \right] =
\delta (\phi'(x)) \;.
\label{A.1}
\end{equation}

When the operator $\hat{\cal O}$ depends only of $\hat{\Phi}$ but not
$\hat{\Pi}$, one easily gets Eq. (\ref{2.7}) since the states $\vert \Phi
\rangle$ are by definition eigenstates of $\hat{\Phi}$ and consequently
${\cal O}(\hat{\Phi}) \vert \Phi \rangle =  {\cal O}(\Phi ) \vert \Phi
\rangle $. It is assumed that ${\cal O}(\hat{\Phi})$ can be expanded in
the power series of $\hat{\Phi}$.

Let us now consider $\hat{\cal O}$ which depends only on $\hat{\Pi}$. Then,
we introduce the complete set of the momentum eigenstates and the r.h.s of
Eq. (\ref{2.7}) is
\begin{eqnarray}
{1\over Z}\; \int {\cal D}\Phi {\cal D}\phi'
{{\cal D}\Pi  \over 2\pi} {{\cal D}\Pi_1\over 2\pi}
{{\cal D}\Pi_2 \over 2\pi} &\; &{\cal O}(\Pi_2 ) \;
{\rm exp}\big[ -i \int dx \;\Pi (x) \phi'(x) \big] \; \nonumber \\
&\times & \langle \Phi + {1 \over 2} \phi' \vert \Pi_1\rangle
\langle \Pi_1\vert \; \hat{\rho}(t) \; \vert \Pi_2 \rangle
\langle \Pi_2\vert \Phi - {1 \over 2} \phi' \rangle \;.
\label{A.2}
\end{eqnarray}
Further we use the momentum eigenfunction
\begin{equation}
\langle \Phi \vert \Pi \rangle =
{\rm exp}\Big[ i\int dx \; \Pi (x) \Phi (x) \Big]
\label{A.3}
\end{equation}
and observe that
\begin{eqnarray}
&{\cal O} & \big( \Pi_2 (x)\big) \;
{\rm exp} \left[ i \int dx \;\Big( \Big[ \Pi_1-\Pi_2 \Big] \Phi
 + {1\over 2} \Big[ \Pi_1+\Pi_2 \Big] \phi'\Big) \right] = \nonumber \\
&\quad &{\cal O}\left( {1 \over i}{\delta \over \delta \phi'(x)}
 -{1 \over 2i} {\delta \over \delta \Phi (x)}\right) \;
{\rm exp}\left[ i \int dx \;\Big( \Big[ \Pi_1-\Pi_2 \Big] \Phi
 + {1\over 2} \Big[ \Pi_1+\Pi_2 \Big] \phi'\Big) \right] \;.
\label{A.4}
\end{eqnarray}
Performing repeatedly the partial integration
we effectively convert ${\cal O}\Big( {1 \over i}{\delta \over \delta
\phi'(x)}   -{1 \over 2i} {\delta \over \delta \Phi (x)}\Big) \;$
into  ${\cal O}\big( \Pi (x)\big)$ and finally prove Eq. (\ref{2.7}) for
${\cal O}\big(\hat{\Pi} (x)\big)$.

When the operator $\hat{\cal O}$ depends simultaneously on
$\hat{\Phi}$ and $\hat{\Pi}$, the situation gets somewhat complicated
because the operators
$\hat{\Phi}$ and $\hat{\Pi}$ do not commute with each other.  The direct
computation, in particular, shows that
\begin{equation}
\langle \hat{\Pi}(x) \hat{\Phi}(y) \rangle =
\langle \Pi (x) \Phi (y) \rangle - {i \over 2}\;\delta (x-y) \;,
\label{A.5}
\end{equation}
while
\begin{equation}
\langle \hat{\Phi}(y) \hat{\Pi}(x) \rangle =
\langle \Pi (x) \Phi (y) \rangle + {i \over 2}\;\delta (x-y) \;.
\label{A.6}
\end{equation}
Therefore,
\begin{equation}
\langle \{ \hat{\Pi}(x), \hat{\Phi}(y) \} \rangle =
\langle \Pi (x) \Phi (y) \rangle \;,
\label{A.7}
\end{equation}
where $\{ \hat{A}, \hat{B} \} \equiv {1 \over 2} \; \Big(
 \hat{A} \hat{B} + \hat{B} \hat{A} \Big)$.

Generalizing the result (\ref{A.7}), one proves the equality
$\langle {\cal O}(\hat{\Phi},\hat{\Pi}) \rangle =
\langle {\cal O}(\Phi ,\Pi ) \rangle $ assuming that
the pairs of noncommuting operators are symmetrized.

\section{}

We derive here the equation of motion (3.3). For this purpose
the Hamiltonian (\ref{3.2}) is split as
\begin{equation}
\hat H = \hat{H}_{\Pi} + \hat{H}_{\nabla} + \hat{H}_m + \hat{H}_I \;,
\label{B.1}
\end{equation}
where $\hat{H}_{\Pi},\; \hat{H}_{\nabla},\; \hat{H}_m ,$ and $\hat{H}_I$
correspond to the first, second, third and fourth term, respectively,
on r.h.s of Eq. (\ref{3.2}).  Now one has to calculate four expressions
\begin{equation}
G_i \equiv \int {\cal D}\phi'\;
{\rm exp}\left[ -{i\over\hbar}\int dx \;\Pi (x) \phi'(x) \right] \;
\langle \Phi + {1 \over 2} \phi' \vert \; [\hat{H}_\alpha ,\hat{\rho}] \;
\vert \Phi - {1 \over 2} \phi' \rangle \;,
\label{B.2}
\end{equation}
with the index $\alpha = \Pi ,\; \nabla ,\; m$ and $I$.

The evaluation of $G_m$ is straightforward since $\vert \Phi \rangle$ is,
by definition, the eigenstate of $\hat{\Phi}$. Thus,
\begin{eqnarray}
G_m &= &{m^2\over 2} \int dx \int {\cal D}\phi' \;
{\rm exp}\left[ -{i\over\hbar}\int dx \;\Pi (x) \phi'(x) \right] \;
\nonumber \\
&\times & \Big( \big(\Phi (x) + {1\over 2} \phi'(x) \big) ^2
- \big(\Phi (x) - {\textstyle{1\over 2}} \phi'(x) \big) ^2 \Big)
\left\langle \Phi  + {\textstyle{1 \over 2}} \phi' \vert \; \hat{\rho} \;
\vert  \Phi - {\textstyle{1 \over 2}} \phi'\right\rangle \;.
\label{B.3}
\end{eqnarray}
Since,
\begin{equation}
\phi'(x)\; {\rm exp}\left[ -{i\over\hbar}\int dx \;\Pi (x) \phi'(x) \right]
= i\hbar {\delta \over \delta \Pi (x)} \;
{\rm exp}\left[ -{i\over\hbar}\int dx \;\Pi (x) \phi'(x) \right] \;,
\label{B.4}
\end{equation}
one finds
\begin{equation}
G_m = i\hbar m^2 \int dx \; \Phi (x) \;{\delta \over \delta \Pi (x)} \;
W[\Phi,\Pi;t] \;.
\label{B.5}
\end{equation}
Keeping in mind that $\hat{\cal L}_I$ is a polynomial in $\hat{\Phi}$,
one also easily computes $G_I$ as
\begin{equation}
G_I = - \int dx \; \left(
{\cal L}_I\Big(\Phi (x) +{i\hbar\over 2}{\delta \over \delta \Pi (x)}\Big) -
{\cal L}_I\Big(\Phi (x) -{i\hbar\over 2}{\delta \over \delta \Pi (x)}\Big)
\right) W[\Phi,\Pi;t] \;.
\label{B.6}
\end{equation}
To find $G_{\Pi}$ we introduce the complete set of the momentum eigenstates
$\vert \Pi \rangle$ and then
\begin{eqnarray}
G_{\Pi} &= &{1\over 2}\int dx \int {\cal D}\phi'
{{\cal D}\Pi_1\over 2\pi} {{\cal D}\Pi_2 \over 2\pi} \;
\exp \left[ -{i\over\hbar}\int dx \;\Pi (x) \phi'(x) \right] \;
\big( \Pi^2_1(x) - \Pi^2_2(x) \big)\nonumber \\
\phantom{=} &\times &\left\langle \Phi + {1 \over 2} \phi'\vert \Pi_1
\right\rangle  \left\langle \Pi_1 \vert \;\hat{\rho}\;\vert \Pi_2 \rangle
\langle \Pi_2 \vert \Phi - {1 \over 2} \phi'\right\rangle \;.
\label{B.7}
\end{eqnarray}

Next we make use of the explicit form of the momentum eigenfunctions
(\ref{A.3}) and observe that
\begin{eqnarray}
{2\hbar^2 \over i^2} {\delta \over \delta \Phi (x)} \;
{\delta \over \delta \phi' (x)} & \; &
{\rm exp}\left[ {i\over\hbar}\int dx \;\Big[ ( \Pi_1 - \Pi_2 )\Phi
+{1\over 2} ( \Pi_1 + \Pi_2 )\phi'  \Big] \right] \nonumber \\
& = & ( \Pi^2_1 - \Pi^2_2 )\;
{\rm exp}\left[ {i\over\hbar}\int dx \;\Big[ ( \Pi_1 - \Pi_2 )\Phi
+{1\over 2} (\Pi_1 + \Pi_2) \phi' \Big] \right] \;.
\label{B.8}
\end{eqnarray}
Since the derivative over $\phi'$ appears under the integral over
$\phi'$, we perform the partial integration and as a result we obtain
\begin{equation}
G_{\Pi} = -i\hbar \int dx \; \Pi (x) \;{\delta \over \delta \Phi (x)} \;
W[\Phi,\Pi;t] \;.
\label{B.9}
\end{equation}
We best find $G_{\nabla}$ defining the field $\Phi (x)$ on a lattice.
The Hamiltonian is then
\begin{equation}
\hat{H}_{\nabla} = {a \over 2} \sum_i \left(
 {\hat{\Phi}_{i+1} - \hat{\Phi}_i \over a} \right) ^2 \;,
\label{B.10}
\end{equation}
where $a$ is the lattice spacing and $i$ numerates the lattice sites.
With such a Hamiltonian the computation of $G_{\nabla}$ is very similar to that
one of $G_m$ and one finds
\begin{equation}
G_{\nabla} = i\hbar a \sum_i{1 \over a^2} (\Phi _{i+1} - \Phi _i) \;
\left( {\partial \over \partial \Pi _{i+1}}
- {\partial \over \partial \Pi _i} \right)
W(\{\Phi_j\} ,\{\Pi_j\} ;t) \;.
\label{B.11}
\end{equation}
In the continuum limit the above expression reads
\begin{equation}
G_{\nabla} = i\hbar \int dx \; \nabla \Phi (x) \;\nabla
{\delta \over \delta \Pi (x)} \; W[\Phi,\Pi;t] \;.
\label{B.12}
\end{equation}
After the partial integration we finally get
\begin{equation}
G_{\nabla} = -i\hbar \int dx \; \nabla^2\Phi (x) \;
{\delta \over \delta \Pi (x)} \; W[\Phi,\Pi;t] \;.
\label{B.13}
\end{equation}
Collecting $G_{\Pi}$, $G_{\nabla}$, $G_m$ and $G_I$ we find the desired
equation
of motion (\ref{3.3}).

\section{}

We derive here the equilibrium Wigner functional of the fields in the
momentum space.  We introduce the complete set of the energy eigenstates
of the two-dimensional (isotropic) oscillator $\vert n_1,n_2 \rangle$
and rewrite Eq. (\ref{2.4}) as
\begin{eqnarray}
\widetilde W[\Phi,\Pi] &=  &\sum_{n_1,n_2} \int  {\cal D}\phi' \;
{\rm exp}\left[ -\int_0^{\infty} dp \Big[ i\big( \Pi ^*(p) \phi'(p)
   + \Pi (p) \Phi^{\prime *}(p) \big)  +\beta E_{n_1,n_2}(p)\Big] \right]
\nonumber \\
\qquad & \times & \langle \Phi + {1 \over 2} \phi' \vert n_1,n_2 \rangle
\langle n_2,n_1 \vert \Phi - {1 \over 2} \phi' \rangle \;,
\label{C.1}
\end{eqnarray}
with $E_{n_1,n_2}(p) \equiv \sqrt{p^2 + m^2}\;(n_1+n_2 + 1)$.
The eigenfunctions for the two dimensional (isotropic) oscillator are
\cite{17}
\begin{eqnarray}
\langle \Phi _R , \Phi _I  \vert n_1,n_2 \rangle & = &\left(
{\alpha ^2 \over  \pi \; 2^{n_1} \; n_1! \; 2^{n_2} \; n_2!}\right)^{1/2}
\nonumber \\
& \times & H_{n_1}(\alpha \Phi _R) \; H_{n_2}(\alpha \Phi _I) \;
 {\rm exp}\left[-{\alpha^2 \over 2}\Big( \Phi_R^2 + \Phi_I^2 \Big)\right] \;,
\label{C.2}
\end{eqnarray}
where $\Phi _{R,I}$ is the real or imaginary part of the field, respectively.
$H_n$ denotes the Hermite polynomial, while
\begin{equation}
\alpha \equiv \Big( {p^2 + m^2 \over 4} \Big) ^{1/4} \;.
\label{C.3}
\end{equation}
Substituting (\ref{C.2}) into (\ref{C.1}) and using the identity \cite{15}
\begin{equation}
\sum_n^{\infty} {a^n \over n!} \; H_n(x) H_n(y) = {1 \over \sqrt{1 - 4a^2}} \;
{\rm exp}\left[ {4axy - 4t^2(x^2 + y^2) \over 1 - 4a^2} \right] \;
\label{C.4}
\end{equation}
which holds for $a<1/2$, one finds after elementary integration and
proper normalization the final
formula of the equilibrium Wigner functional (\ref{4.3}).
The same result can be obtained by analytic continuation and generalization
to infinitely many degrees of freedom of the formula given in the Appendix
of ref.\cite{14c} for the generating function $G(s)$ for the Wigner functions
associated with the energy eigenstates of the harmonic oscillator.

\section{}

The purpose of this Appendix is to derive the generating functional
(\ref{5.3}).  We apply the method described in the literature \cite{18}
modifying it slightly.  After integration over $\Pi$ and partial
integrations with respect to $x'$ and then $x$, we rewrite Eq. (\ref{5.1}) as
\begin{equation}
{\cal Z}[j] = {\cal C} \int {\cal D}\Phi \;
 {\rm exp} \left[ - \beta \int dx \;dx' \Big[{1 \over 2} \Delta (x - x')
 \Phi (x) \big( m^2- \nabla^2 \big) \Phi (x')
 - \delta (x- x') \Phi (x) j(x') \Big] \right]\;.
\label{D.1}
\end{equation}
It is important for these manipulations that $\Delta (x) = \Delta (-x)$.
Now we change the variable $\Phi (x)$ into $\Phi (x) + \Phi_0(x)$ demanding
that the field $\Phi_0(x)$ satisfies the equation
\begin{equation}
\int dx' \Delta (x-x')\;\big( -\nabla^2 + m^2 \big) \; \Phi_0(x')
= j(x) \;.
\label{D.2}
\end{equation}
After integration over $\Phi$ and the partial integrations with respect to $x$
and $x'$, we get the result
\begin{equation}
{\cal Z}[j] = {\cal N}\;  {\rm exp} \left[
{\beta \over 2} \int dx \;   \Phi_0 (x) j(x) \right] \;.
\label{D.3}
\end{equation}
Substituting into Eq. (\ref{D.3}) the solution of Eq. (\ref{D.2}) in the form
\begin{equation}
\Phi_0(x) = \int dx' \; {\cal G}(x-x') j(x') \;,
\label{D.4}
\end{equation}
with ${\cal G}$ being the Green function given by Eq. (\ref{5.4}),
one finally finds the generating functional (\ref{5.3}).

\end{document}